**Unconventional exchange bias coupling at perovskite/brownmillerite interface in spontaneously stabilized $SrCoO_{3-\delta}$/$SrCoO_{2.5}$ bi-layer**

B. C. Behera[1], Subhadip Jana[1], Shwetha G. Bhat[2], G. Tripathy[1, 3], P. S. Anil Kumar[2], D. Samal[*1, 3]

[1]Institute of Physics, Sachivalaya Marg, Bhubaneswar – 751005, India

[2]Department of Physics, Indian Institute of Science Bangalore – 560012, India

[3]Homi Bhabha National Institute, AnushaktiNagar, Mumbai – 400085, India

Interface effect in complex oxide thin film heterostructures lies at the vanguard of current research to design technologically relevant functionality and explore emergent physical phenomena. While most of the previous works focus on the perovskite/perovskite heterostructures, the study on perovskite/brownmillerite interfaces remain at its infancy. Here, we investigate spontaneously stabilized perovskite-ferromagnet ($SrCoO_{3-\delta}$)/brownmillertite-antiferromagnet ($SrCoO_{2.5}$) bi-layer with $T_N > T_C$ and discover an unconventional interfacial magnetic exchange bias effect. From magnetometry investigations, it is rationalized that the observed effect stems from the interfacial ferromagnet/antiferromagnet coupling. The possibility for coupled ferromagnet/spinglass interface engendering such effect is ruled out. Strikingly, a finite coercive field persists in the paramagnetic state of $SrCoO_{3-\delta}$ whereas the exchange bias field vanishes at $T_C$. We conjecture the observed effect to be due to the effective external quenched staggered field provided by the antiferromagtic layer for the ferromagnetic spins at the interface. Our results not only unveil a new paradigm to tailor the interfacial magnetic properties in oxide heterostructures without altering the cations at the interface, but also provide a purview to delve into the fundamental aspects of exchange bias in such unusual systems paving a big step forward in thin film magnetism.

The investigation on interfacial magnetic effects in transition metal oxide based thin film heterostructures has sparked unprecedented scientific developments and is pursued intensively because of its technological promise for the next-generation nano-scaled magnetic devices [1]. A precise control and tuning of interfacial magnetic properties in thin film heterostructures is crucial for engendering exotic functionalities which are highly relevant for technological applications such as magnetic field sensors, memories or magnetic recording read heads [2-4]. A great deal of attention in this regard is focused on the effect called "exchange bias" that occurs due to interfacial magnetic exchange coupling in a coupled ferromagnetic/antiferromagnetic system [5, 6]. This effect is widely maneuvered for the design and operation of spin valve based magnetic read heads and sensors. The macroscopic hallmark of magnetic exchange bias effect (MEBE) is the unidirectional shift of the $M$(H) loop along the field-axis (**Figure**1(c)), and enhancement of coercivity. Typically, a bi-layer consisting of a FM and an AF (with the Curie temperature ($T_C$) of FM greater than the Néel temperature ($T_N$) of AF) when cooled in a static magnetic field across the $T_N$, an unidirectional exchange anisotropy-fieldgets locked in and give rise to exchange bias effectthat stabilizes the orientation of the ferromagnetic layer [2, 7, 8]. Such systems exhibit magnetic properties that markedly differ from their constituents. Though exchange bias related phenomena in FM/AFM coupled system is studied extensively, its inherent mechanism has not been completely understood because it is always hard nut to directly observe and manipulate the spin structure at the interface.

Hitherto MEBE has been observed in numerous metal-oxides FM/AF coupled systems (e.g. $La_{0.67}Sr_{0.37}MnO_3/SrMnO_3$ [9], $La_{0.67}Ca_{0.37}MnO_3/SrMnO_3$ [10], $La_{0.67}Sr_{0.37}MnO_3/BiFeO_3$ [11], $La_{0.67}Sr_{0.37}MnO_3/TbMnO_3$ [12] with the $T_N$ of the accompanying AF being always lower than the $T_C$ of the FM. Thus, it has been generally accepted that $T_C > T_N$ criterion is a prerequisite for establishing exchange bias effect at the FM/AF interface. Indeed, all theoretical models have virtually relied on the assumption of $T_C > T_N$ criterion to delve into the mechanism for interfacial coupling [5, 6, 13, 14]. Regardless of long orthodox belief of $T_C > T_N$ criterion, a few remarkable experimental observations manifesting MEBE are reported in FM/AF bi-layer systems with $T_N > T_C$. For instance, Chein *et al.* observed exchange coupling phenomena in FM/AF bi-layers of $a$-(Fe$_{0.1}$Ni$_{0.9}$)$_{80}$B$_{20}$ ($T_C \sim 240$ K)/CoO ($T_N = 291$ K) and $a$-Fe$_4$Ni$_{76}$B$_{20}$ ($T_C \sim 150$ K)/CoO ($T_N \sim 291$ K), respectively [15, 16]. Similar effect was also observed in the MnO (antiferromagnet)/Mn$_3$O$_4$ (ferrimagnet) core/shell structure [17]. Here we explore such effect in a perovskite(SrCoO$_{3-\delta}$)/brownmillerite (SrCoO$_{2.5}$) thin-film interface; wherein bulk SrCoO$_3$ hosts a metallic ferromagnetic state with $T_C$ in the range 280-305 K [18, 19]and SrCoO$_{2.5}$ exhibits an insulating antiferromagnetic state with $T_N \sim 570$ K [20, 21] as sketched in **Figure**1(a)).

SrCoO$_x$ exhibits highly contrastingstructural, electronic, and magnetic property depending on the Co oxidation state, which can be manipulated by controlling the oxygen stoichiometry [22-27]. Brownmillerite SrCoO$_{2.5}$ (SCO$_{BM}$) derives its structure from the perovskite SrCoO$_3$ (SCO$_{PC}$) through the removal of 1/6$^{th}$ of oxygen atoms such that alternating oxygen octahedral and tetrahedral are stacked together [18]. While bulk SCO$_{BM}$ is readily synthesized under ambient condition, SCO$_{PC}$ limits its synthesis to extremely high pressure, due to relatively large thermodynamic energy barrier for the formation of perovskite phase involving Co$^{4+}$ ions. However, recent studies reveal the epitaxial stabilization of single crystalline SCO$_{PC}$ thin films *via* topotactic phase transformation under high oxidizing condition [23-25]. Manipulating the oxygen sublattice in complex oxide thin filmheterostructure/interface offers a promising avenue to look for fascinating functionality. Here we report the fabrication of FM-perovskite(SrCoO$_{3-\delta}$)/AFM-brownmillerite (SrCoO$_{2.5}$) natural bi-layer by pulsed laser epitaxy and demonstrate the evidence forunconventional exchange biaswith $T_N > T_C$. The term natural bi-layer is coined categorically to emphasize that an interface involving perovskite-SrCoO$_3$ and brownmillertite-SrCoO$_{2.5}$ is formed spontaneously as shown in **Figure.**1 (b).The perovskite/brownmillerite interfaces are expected to host emergent interfacial phenomena due to lattice symmetry mismatch, but remains scantly explored [28, 29]. A recent study by Zhang *et al.* found robust perpendicular magnetic anisotropy for $La_{2/3}Sr_{1/3}MnO_3/LaCoO_{2.5}$ [28]. The present revelation for the spontaneous stabilization of perovskite/brownmillerite ($SCO_{PC}/SCO_{BM}$) interface without altering the cations uniquely provides an ideal system to investigate the interfacial phenomena. In particular, the realization of unusual MEBE in (FM-$SCO_{PC}$/AFM-$SCO_{BM}$) bi-layer in this study constitutes a fundamental step to broaden the search to a greater variety of FM/AF bi-layer with $T_N > T_C$ exhibiting exchange bias related phenomena that will have relevant implications in technological applications.

High-quality epitaxial natural bi-layers consisting of $SCO_{PC}$ and $SCO_{BM}$ with varying thickness were grown on the single-crystalline STO(001) substrates ($a$ = 3.905 Å) using pulsed laser deposition. $SCO_{PC}$ exhibits a typical $ABO_3$ perovskite structure with cubic $Pm\underline{3}m$ symmetry ($a$ = 3.829 Å) while $SCO_{BM}$ forms an orthorhombic *Ima2* symmetry with $a$ = 5.574 Å, $b$ = 5.469 Å and $c$ = 15.745 Å [30]. In a pseudo-tetragonal setting, the lattice parameters of $SCO_{BM}$ can be expressed as $a_t$ = 3.905 Å and $c_t$ = 3.936 Å, where, $a_t$ and $c_t$ are the lattice parameter of the reduced pseudo-tetragonal representation of *Ima2* symmetry. We present the results on two representative bi-layers: [$SCO_{PC}$(~ 24nm)/$SCO_{BM}$(~ 2nm)] and [$SCO_{PC}$(~ 20nm)/$SCO_{BM}$(~ 2nm)]. The structural characterization is carried out using high-resolution X-ray diffractometer (Rigaku, Smart Lab). The *θ-2θ* XRD pattern of bi-layers is found to be oriented along (*00l*) as shown in **Figure** 1(d). The full range of *θ-2θ* XRD pattern is shown in the SM (**Figure** S1). To determine the thickness of individual layers and the stacking order, X-ray reflectivity (XRR) measurement is performed and the corresponding data was fitted using Globalfit software of Rigaku (**Figure** 1(f) and **Figure**S2 (SM)). The results from XRR fitting inferred the stacking order to follow [STO/$SCO_{PC}$/$SCO_{BM}$] type, indicating the possible oxygen vacancies in proximity to surface could drive the top layer into BM phase in the bi-layers. The reverse order of stacking such as [STO/ $SCO_{BM}/SCO_{PC}$] was ruled out as the latter gave an inadequate fit to the XRR data. From XRR fitting, the average surface roughness of the bi-layer was found to be ~ 0.5 nm, indicating an atomically smooth surface. The observation of Kiessig fringes and well-pronounced Laue oscillations in the x-ray spectra are also a clear indication of superior quality of the bi-layers. The out-of-plane lattice parameters "*c*" calculated from (*00l*) peak positions for [$SCO_{PC}$ (~ 24nm)/$SCO_{BM}$ (~ 2nm)] bi-layer are 3.809 Å and 3.920 Å for $SCO_{PC}$ and $SCO_{BM}$ respectively. These values are comparatively smaller than the lattice parameter for their bulk counterparts, suggesting an out-of-plane compressive stress. Similar observations are also observed for the other [$SCO_{PC}$ (~ 20nm)/$SCO_{BM}$ (~ 2nm)] bi-layer (see SM). To elucidate the in-plane epitaxial relationship in the bi-layers, we performed the φ-scan and reciprocal space mapping with respect to STO (103), $SCO_{PC}$ (103) and $SCO_{BM}$ (11$\underline{1}$2) planes. Four equally spaced distinct peaks with a relative separation of 90º (four-fold symmetry) were observed in the φ-scans, suggesting the cube on cube epitaxial growth of the constituent layers on STO i.e:[100]STO∥[100]$SCO_{PC}$∥[100]$SCO_{BM}$. From reciprocal mapping, it was observed that the diffraction peaks associated with STO and the constituting layers lie at the same $q_x$ value, and thus indicates the bi-layer to be completely strained with respect to the underlying substrate (**Figure** 1(g)). In essence, our extensive structural investigation elucidates the occurrence of strained epitaxial $SrCoO_{3-\delta}/SrCoO_{2.5}$ natural bi-layer on STO.

To characterize the magnetic properties of the $SrCoO_{3-\delta}/SrCoO_{2.5}$ bi-layer films, we have measured magnetization (*M*) versus temperature (T) and magnetization (M) versus field (H) in zero field cooled (ZFC) and field cooled (FC) mode. Systematic analysis of the magnetic data after correcting for the diamagnetic substrate contribution was carried out to determine the coercive and exchange bias field (see SM). In **Figure** 2 (a), we show the FC and ZFC temperature dependent magnetization $M$(T) for [SCO$_{PC}$(~24nm)/SCO$_{BM}$ (~2nm)] bi-layer (hereafter abbreviated as (SCO$_{PC}$/SCO$_{BM}$) for the sake of brevity unless explicitly clarified) with a field of 100 Oe applied along the in-plane direction of (001) STO. A ferromagnetic like order is apparent from $M$(T) with an onset of transition at ~175 K. The value of observed $T_C$ ~175K is found to be lower than that of its SCO$_{PC}$ bulk counterpart in which the $T_C$ ranges from 280-305 K [18, 19]. The diminished $T_C$ in thin films could be attributed to finite size $(T_C(\infty)-T_C(t))/T_C(\infty)=(c/t)^\lambda$, where ($T_c(\infty)$) is the curie temperature in the bulk limit, $T_c(t)$ is the curie temperature of the films according to their thickness, $c$ is related to spin correlation length, $t$ is the film thickness and $\lambda$ is the critical shift exponent) and strain effects [23, 24, 31-33]. Indeed, it has been reported that as the degree of substrate induced tensile stain increases from 0.9% ((LaAlO$_3$)$_{0.3}$-(SrAl$_{0.5}$Ta$_{0.5}$O$_3$)$_{0.7}$)) to 1.8% (STO), the ferromagnetic transition temperature reduces from ~250 K to ~200 K [23, 24]. Besides, we observe a finite magnetization above $T_C$ unlike a conventional FM-paramagnetic (PM) transition (**Figure 2(a)**). Similar features are also observed in the [SCO$_{PC}$ (~ 20nm)/SCO$_{BM}$ (~ 2nm)] bi-layer (see the SM). The finite magnetization above $T_C$ in both the bi-layers could be attributed to the contribution from the antiferromagnetic SCO$_{BM}$ layer. Albeit the direct probe to AF transition SrCoO$_{3-\delta}$/SrCoO$_{2.5}$ bi-layers is constrained because of experimental limitation to access high temperature magnetic measurements, sufficient evidences based on magnetometery measurements are inferred in the succeeding section to warrant the existence of antiferromagnetic SCO$_{BM}$ layer conjointly with ferromagnetic SCO$_{PC}$ layer. To substantiate about the existence of antiferromagnetic character in (SCO$_{PC}$/SCO$_{BM}$) bi-layer, we present the ZFC$M$(H) plot at 5 K. Few noteworthy points can be identified from the $M$(H) loop: (a) it shows a ferromagnetic-like hysteresis loop, however the loop gets significantly constricted near zero field region (b) magnetization increases monotonically with applied magnetic field with no saturation even up to 5 Tesla (c) the magnetization value obtained at 5 Tesla is about 1 $\mu_B$/u.c., which is significantly smaller than values reported for the single crystal (~ 2.5 $\mu_B/Co^{4+}$ [18]) and epitaxial thin films of only $SCO_{PC}$ [23]. All the above observations in the ZFC $M(H)$ loop indicate towards the possible existence of FM/AF exchange coupling at the interface in ($SCO_{PC}/SCO_{BM}$) bi-layer.

The interfacial FM/AF exchange coupling is widely probed by MEBE thatshowsadisplacement of the ferromagnetic hysteresis loop along the magnetic field axis.Further, the direction of the loop-shift reverses when the cooling field is reversed. Followed by the indication for interfacial magnetic exchange coupling from the ZFC $M(H)$ plots, we measured the field cooled $M(H)$ loop to examine the possible emergence of MEBE. In-plane magnetic hysteresis loops measured at 5 K after field cooling from 350 to 5 K in applied fields of +3 T and –3 T are shown in **Figure** 2(c). Interestingly, the center of magnetic hysteresis loop (MHL) was observed to shift by ~ 105 Oe along the -ve and +ve side of field axes for applied fields of +3 T and –3 T, respectively, manifesting the sign of negative MEBE. The observed effect in the present case is reminiscent of the recent work by A. Miglioorini *et al*. that reported the spontaneous exchange bias formation driven by a structural phase transition associated with IrMn in IrMn/FeCo bi-layer [34]. To shed light on the possible microscopic origin of the observed MEBE in the present case, we consider two possible structures as shown schematically in **Figure** 2(d). In one case, we consider layered by layered structure with a sharp FM/AF interface, and in other case we consider a random mixture of FM and AF clusters that could give rise to a spin glass (SG) like phase at the interfaces due to inter-cluster interaction. Though the manifestation of MEBE is commonly observed when a FM is in contact with an AF, there are instances of observing such effect in FM/SG coupled systems [35]. If the observed MEBE was due to FM-SG coupling, it would be natural to expect time dependent slow dynamics response in magnetization since SG state are intrinsic to numerous meta-stable states. In a SG system, time decay of the thermo-remnant magnetization (TRM) generally follows a logarithmic trend [36]. As shown in **Figure** 2(e), the TRM for (SCO$_{PC}$/SCO$_{BM}$) bi-layer in contrast remains almost constant over four decades of observation time (see SM for the protocol used for TRM measurement). Thus, we rule out the possibility of any coupled FM/SG interface resulting in MEBE in (SCO$_{PC}$/SCO$_{BM}$) bi-layer.

Further, to elucidate about the temperature dependence of MEBE, we measured field cooled *M*(H) loops at various temperatures (**Figure** S3 in SM). The measured loops revealed a systematic change in asymmetry and coercivity. For every successive measurement at each fixed temperature, the sample was field cooled from 350 K with an applied field of 3T. To demonstrate the MEBE more clearly, we have plotted the MHL using the inversion method [37], in which M and H of the reversing part of the original loop are multiplied by -1 and the modified loop so called "inverted loop" is presented in **Figure** 3. In inverted loop, the $H_{C-}$ value of the original loop gets shifted to positive field side, and thus we can see the difference between $H_{C+}$ of original loop and $H_{C-}$ of inverted loop with clarity. Enlarged curves of original and inverted hysteresis loops are shown in **Figures** 3(a)-(i). It is evident that MEBE gradually becomes weaker with increasing temperature. The characteristic exchange bias field $H_{EB}$ and the coercive field $H_C$ are estimated using the relations $H_{EB} = (H_{C+} + H_{C-})/2$ and $H_C = (H_{C+} - H_{C-})/2$, where $H_{C+}$ and $H_{C-}$ are coercive fields for the positive and negative field axes of original *M*(H) loop, respectively as mentioned earlier. **Figures** 3(j) and 3(k) summaries the temperature dependence of $H_{EB}$ and $H_C$ in which a maximum $H_{EB}$ and $H_C$ of 105 and 450 Oe respectively is observed at 5 K. Remarkably, it has to be noted that while $H_{EB}$ gradually falls to zero at $T_C$ ~ 175 K, a striking non-zero coercive field $H_C$ persists beyond $T_C$. Earlier studies on FM/AF (FeNiB/CoO) bi-layer with $T_N$ (291 K) > $T_C$ (~150 K) reported the persistence of $H_{EB}$ well above $T_C$ *i.e.* into the paramagnetic state of the ferromagnet though $H_C$ fell to zero at $T_C$ [16]. The induced $H_{EB}$ in the paramagnetic state in case of FeNiB/CoO bi-layer was attributed to modest magnetization in the paramagnetic state of FeNiB, originated either from field cooling effect or some local ordering at the interface due to close proximity to AF layer. Unlike FeNiB/CoO bi-layer case, in the present study we do not see any signature MEBE effect indicating that no net interfacial exchange bias persists above $T_C$. However the observation of a finite $H_C$ for $T > T_C$ is intriguing. We conjecture that this counterintuitive observation results from the exchange coupling of the magnetic moments in the FM layer to those in the adjacent AF layer, the latter in this case providing an effective staggered external field for the FM spins. The magnetic moments in the AFM layer are effectively quenched since $T << T_N$. Preliminary Monte Carlo simulations of simple Ising like models of the coupled FM-AF layers with $T_C < T << T_N$ supports this hypothesis (see **Figure** S8 SM).

Finally, we examine about the nature and origin of MEBE in (SCO$_{PC}$/SCO$_{BM}$) bi-layer by performing field cooled MHL measurements at 5 K under varying biasing-field strength as shown in **Figure** 4. It is evident that the asymmetry in MHL widens as the biasing cooling-field strength increases. The estimated $H_{EB}$ under various cooling field strength shows a saturating tendency towards higher field (a maximum $H_{EB}$ of 140Oe is observed at biasing field strength of 5T). Such saturating tendency of $H_{EB}$ at higher biasing field are common in FM/AF coupled systems unlike the case for coupled FM/SG system in which $H_{EB}$ typically gets reduced for large cooling field [17]. This indicates that the observed interfacial coupling in epitaxial (SCO$_{PC}$/SCO$_{BM}$) bi-layer is FM/AF type rather than FM/SG one. The observations made both from varying biasing -field and time dependent magnetization study validate each other.

In summary, we have demonstrated that a hetero-interface involving perovskite-ferromagnet SrCoO$_{3-\delta}$ and brownmillertite-antifferomagnet SrCoO$_{2.5}$ with $T_N > T_C$ is formed spontaneously by the modification of oxygen sublattice using pulsed laser epitaxy and exhibits unusual MEBE. This is contrary to common perception in which $T_C > T_N$ criterion is generally considered to observe MEBE at FM/AF interface. Structural findings testify the occurrence of SrCoO$_{3-\delta}$/SrCoO$_{2.5}$ epitaxial natural bi-layer on STO. Detailed magnetometery investigations reveal the central footprints for FM/AF interfacial exchange coupling. The possibility of coupled FM/SG interfacial coupling giving rise to MEBE is ruled out by time dependent thermo-remnant and biasing-field dependent $H_{EB}$ measurements. Interestingly, we observe a finite coercive field in the paramagnetic state of SrCoO$_{3-\delta}$ whereas the exchange bias field vanishes at $T_C$. We conjecture that this counterintuitive observation is due to the effective external quenched staggered field provided by the AFM spins. In essence, the present workoffers anew perspective to design innovative interfaces between oxides of different structural symmetryand explore emergent interfacial phenomena. Moreover, we believe that the observation of MEBE in (SCO$_{PC}$/SCO$_{BM}$) bi-layer will extend the realm of exchange bias beyond conventional systems and broaden the search to a greater combination of FM/AF bi-layers with $T_N > T_C$ resulting in such effect. The basic understanding of such unusual exchange bias phenomena will trigger a big step forward in thin film and interfacial magnetism.

**Experimental Section**

*Thin film hybrid growth:* Spontaneously stabilized high-quality epitaxial SrCoO$_{3-\delta}$(SCO$_{PC}$)/SrCoO$_{2.5}$ (SCO$_{BM}$) bi-layers of different thickness ([SCO$_{PC}$ (~ 20 nm)/SCO$_{BM}$ (~ 2 nm)] and [SCO$_{PC}$ (~ 24 nm)/SCO$_{BM}$ (~ 2 nm)]) were fabricated on the single-crystalline (001) oriented SrTiO$_3$ (STO) substrates using Pulsed laser deposition (PLD). A KrF excimer laser ($\lambda$ = 248 nm)) with a fluence of 2 Jcm$^{-2}$ and a repetition rate of 2 Hz were adopted to ablate material from polycrystalline SrCoO$_{2.5}$ target. The substrate–target separation was fixed at 60 mm. Before deposition the target was pre-ablated for 2 minutes at a pulse rate of 5 Hz and laser fluence 2 Jcm$^{-2}$. All the bilayer structures were grown at 0.2 mbar of oxygen partial pressure (p(O$_2$)) and a substrate temperature of 750 ºC. After growth, the bilayer structures were cooled down to 400 ºC under same growth pressure and then annealed at 400 ºC for 20 min by introducing p(O$_2$) of 700 mbar. Finally, the temperature was ramped down to room temperature under the same annealing pressure.

*Structural characterization:* The structural characterizations were carried out using high-resolution X-ray diffractometer (Rigaku, Smart Lab). High angle *θ-2θ* X-ray scan, *ϕ*-scan, X-ray reflectivity (XRR), reciprocal space mapping (RSM) were performed to investigate the structural quality of the bi-layers.

*Magnetic measurements:* Magnetic measurements were performed by a superconducting quantum interface device based magnetometer (Quantum Design SQUID-VSM). The magnetic field (H) was set to zero in an oscillation mode to reduce the residual field of the magnet before measurements. The residual field was further calibrated by a reference Pd sample that shows a negligible value (see the SM). TRM measurement was performed using the measurement protocol as mentioned in SM.

**Supporting Information**

Supporting Information is available from the Wiley Online Library or from the author.

**Acknowledgements**

We are grateful to Sachin Sarangi for his superb technical support during magnetic measurements. We thank Gopal Pradhan for fruitful discussion. We thank ZhichengZhong for reading the manuscript and suggestion. We thank T Som for extending laboratory facility. D. Samal and B. C. Behera acknowledge the financial support from Max-Planck Society through Max Planck Partner Group.Shwetha G. Bhat acknowledges INSPIRE Faculty Fellowship Programme for the financial support.

-----------------------------------------------------------

[*]dsamal@iopb.res.in

References

[1] F. Hellman, A. Hoffmann, Y. Tserkovnyak, G. S. D. Beach, E. E. Fullerton, C. Leighton, A. H. MacDonald, D. C. Ralph, D. A. Arena, H. A. Dürr, P. Fischer, J. Grollier, J. P. Heremans, T. Jungwirth, A. V. Kimel, B. Koopmans, I. N. Krivorotov, S. J. May, A. K. Petford-Long, J. M. Rondinelli, N. Samarth, I. K. Schuller, A. N. Slavin, M. D. Stiles, O. Tchernyshyov, A. Thiaville, and B. L. Zink, *Rev. Mod. Phys.* 2017, **89**, 025006.

[2] J.Noguésa, J.Sorta, V.Langlaisb, V.Skumryeva, S.Suriñachb, J.S.Muñozb, M.D.Baró, *Phys. Rep.* 2005, **422**, 65.

[3] M. Bibes, J. E. Villegas, andA. Barthelemy, *Adv. Phys.* 2011, **60**, 5.

[4] C. Chappert, A. Fert, F. N. Van Dau, Nat. Mater.2007, **6**, 813.

[5] W. H. Meiklejohn and C. P. Bean, *Phys. Rev.*1956, **102**, 1413.

[6] W. H. Meiklejohn and C. P. Bean, *Phys. Rev.*1957, **105**, 904.

[7] J. Nogues and I. K. Schuller,J. Magn. Magn. Mater., 1999, **192**, 203.

[8] M. Kiwi, J. Magn. Magn. Mater., 2001, **234**, 584.

[9] J. F. Ding, O. I. Lebedev, S. Turner, Y. F. Tian, W. J. Hu, J. W. Seo, C. Panagopoulos, W. Prellier, G. Van Tendeloo, and T. Wu, *Phys. Rev. B.*2013, **87**,054428.

[10] T. Yu, X. K. Ning, W. Liu, J. N. Feng, X. G. Zhao, and Z. D. Zhang, *J. Appl. Phys.*2014, **116**, 083908.

[11] Y. K. Liu, Y. W. Yin, S. N. Dong, S. W. Yang, T.Jiang, X. G. Li, *Appl. Phys. Lett.*2014, **104**, 043507.

[12] Y. F. Tian, O. I. Lebedev, V. V. Roddatis, W. N. Lin, J. F. Ding, S. J. Hu, S. S. Yan, T. Wu, *Appl. Phys. Lett.*2014, **104**, 152404.

[13] D. Mauri, H. C. Siegmann, P. S. Bagus, and E. Kay, *J. Appl. Phys.***62**, 3047 (1987).

[14] N. C. Koon, *Phys. Rev. Lett.*1997, **78**, 4865.

[15] X. W. Wu and C. L. Chien, *Phys. Rev. Lett.*1998, **81**, 2795.

[16] J. W. Cai, Kai Liu, and C. L. Chien, *Phys. Rev. B.*1999, **60**, 72.

[17] G. S. Alvarez, J. Sort, S. Surinach, M. D. Baro, and J. Nogués, *J. Am. Chem. Soc.* 2007, **129**, 9102.

[18] Y. Long, Y. Kaneko, S. Ishiwata, Y. Taguchiand Y. Tokura, *J. Phys.: Condens. Matter* 2011, **23**, 245601.

[19] C. K. Xie, Y. F. Nie, B. O. Wells, J. I. Budnick, W. A. Hines, and B. Dabrowsk, *Appl. Phys. Lett.* 2011, **99**, 052503.

[20] T. Takeda, Y. Yamaguchi and H. Watanabe, *J. Phys. Soc. Jpn.*, 1972, **33**, 970.

[21] A. Muñoz, C. de la Calle, J. A. Alonso, P. M. Botta, V. Pardo, D. Baldomir, and J. Rivas, *Phys. Rev. B.* 2008, **78**, 054404.

[22] C. K. Xie, Y. F. Nie, B. O. Wells, J. I. Budnick, W. A. Hines, and B. Dabrowsk, *Appl. Phys. Lett.* 2011, **99**, 052503.

[23] H. Jeen, W. S. Choi, M. D. Biegalski, C. M. Folkman, I. C. Tung, D. D. Fong, J. W. Freeland, D. Shin, H. Ohta, M. F. Chisholm, and H. N. Lee, *Nature Mater.* 2013, **12**, 1057.

[24] H. Jeen, W. S. Choi, J. W. Freeland, H. Ohta, C. U. Jung and H. N. Lee, *Adv. Mater.* 2013, **25**, 3651.

[25] W. S. Choi, H. Jeen, J. H. Lee, S. S. Ambrose Seo, V. R. Cooper, K. M. Rabe, and H. N. Lee, *Phys. Rev. Lett.* 2013, **111**, 097401.

[26] N. Lu, P. Zhang, Q. Zhang, R. Qiao, Q. He, Hao-Bo Li, Y. Wang, J. Guo, D. Zhang, Z. Duan, Z. Li, M. Wang, S. Yang, M. Yan, E. Arenholz, S. Zhou, W. Yang, L. Gu, Ce-Wen Nan, J. Wu, Y. Tokura and Pu Yu, *Nature*, 2017, **546**, 124.

[27] B. Cui, C. Song, F. Li, X. Y. Zhong, Z. C. Wang, P. Werner, Y. D. Gu, H. Q. Wu, M. S. Saleem, S. S. P. Parkin, and F. Pan, *Phys. Rev. Applied* 2017, **8**, 044007.

[28] J. Zhang, Z. Zhong, X. Guan, X. Shen, J. Zhang, F. Han, H. Zhang, H. Zhang, X. Yan, Q. Zhang, L. Gu, F. Hu, R. Yu, B. Shen, and J. Sun, *Nat. Commun.* 2018, **9**, 1923.

[29] T. L. Meyer, H. Jeen, X. Gao, J. R. Petrie, M. F. Chisholm, and H. N. Lee, *Adv. Electron. Mater.* 2016, **2**, 1500201.

[30] H. Jeen, and H. N. Lee, *AIP Advances* 2015, **5**, 127123.

[31] M. E. Fisher and M. N. Barber, *Phys. Rev. Lett.* 1972, **28**, 1516.

[32] C. Xiea, J. I. Budnick, and B. O. Wells, *Appl. Phys. Lett.* 2007, **91**, 172509.

[33] J. R. Petrie, C. Mitra, H. Jeen, W. S. Choi, T. L. Meyer, F. A. Reboredo, J. W. Freeland, G. Eres, and H. N. Lee, *Adv. Funct. Mater.* 2016, **26**, 1564.

[34] A. Migliorini, B. Kuerbanjiang, T. Huminiuc, D. Kepaptsoglou, M. Muñoz, JLF Cuñado, J. Camarero, C. Aroca, G .Vallejo-Fernández, VK.Lazarov, JL Prieto, *Nat Mater.* 2018, **17**, 28.

[35] M. Ali, P. Adie, C. H. Marrows, D. Greig, B. J. Hickey, and R. L. Stamps, *Nat Mater.* 2007, **6**,70.

[36] J. A. Mydosh, Spin glasses: An experimental Introduction (Taylor & Francis,London,1993).

[37] Y. Zhou, L. Miao, P. Wang, F. F. Zhu, W. X. Jiang, S. W. Jiang, Y. Zhang, B. Lei, X. H. Chen, H. F. Ding, H. Zheng, W. T. Zhang, J.Jia, D. Qian, and D. Wu1, *Phys. Rev. Lett.* 2018, **120**, 097001.

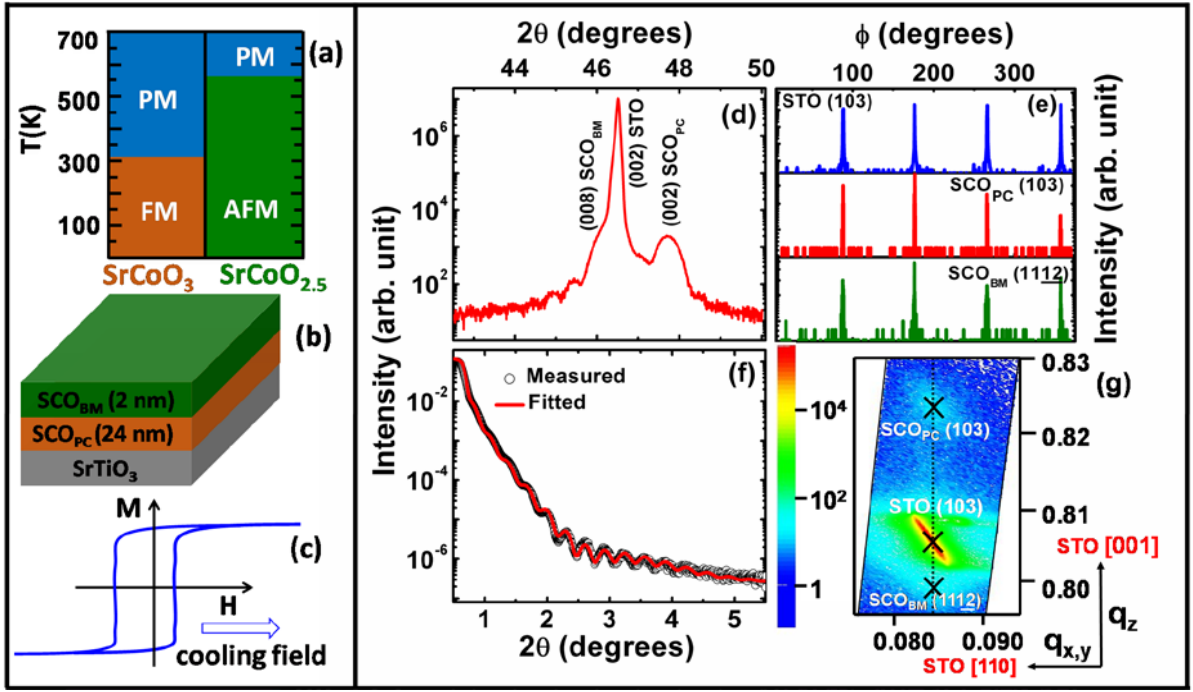

**Figure 1**: left panel **(a)** schematic representation of magnetic ordering temperature for ferromagnetic $SrCoO_3$ and antiferromagnetic $SrCoO_{2.5}$, **(b)** layout of the designed [$SCO_{PC}$(~ 24nm)/$SCO_{BM}$(~ 2nm)] bi-layer on STO, **(c)** schematic MHL representing MEBE under positive field cooling. Right panel **(d)** θ-2θ x-ray diffraction pattern, **(e)** φ-scan along the asymmetric planes of STO(103), $SCO_{PC}$(103), $SCO_{BM}$(11$\underline{1}$2), **(f)** measured and fitted x-ray reflectivity, and **(g)** off-specular reciprocal space mapping around STO(103) of [$SCO_{PC}$(~ 24nm)/$SCO_{BM}$ (~ 2nm)] bi-layer.

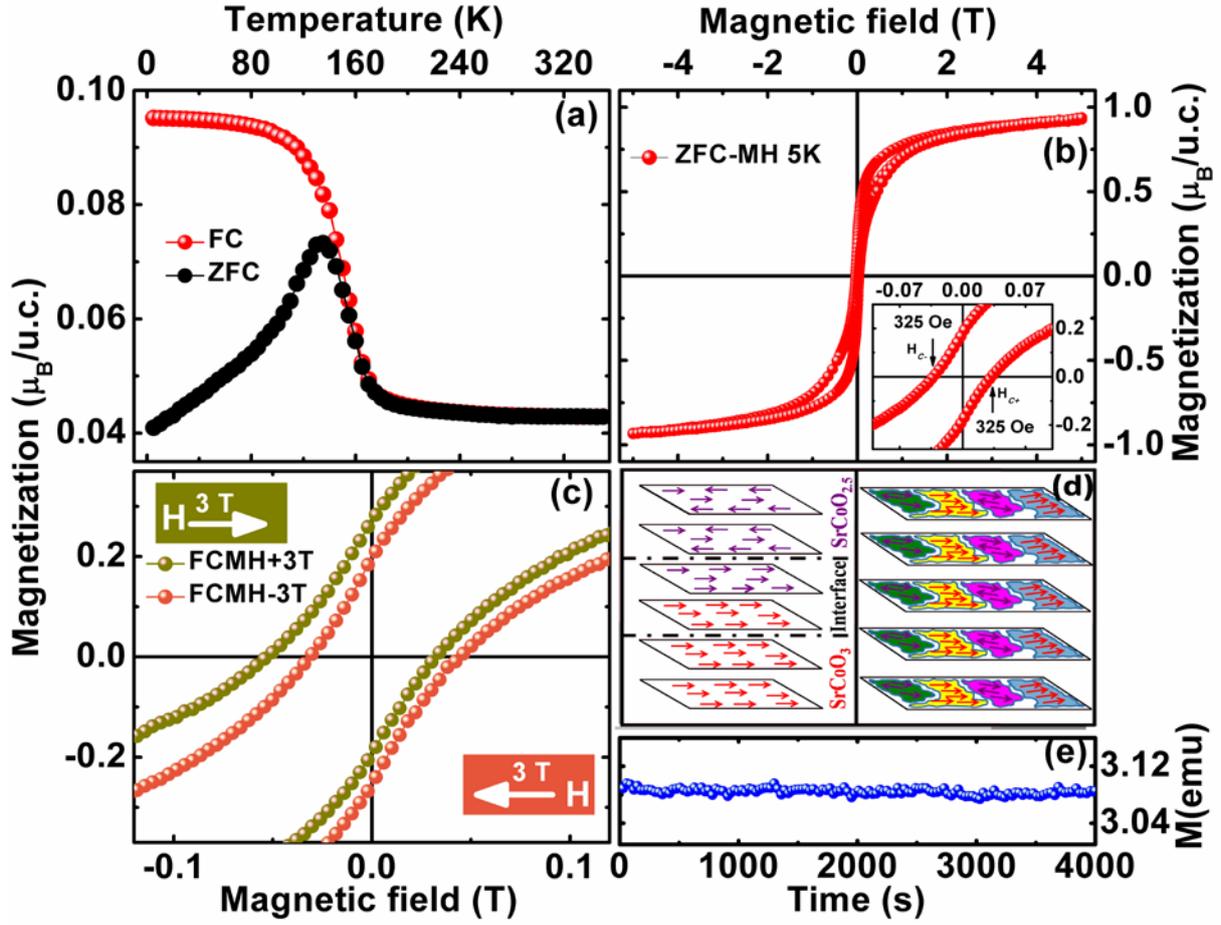

**Figure 2**: **(a)** Temperature dependent zero field cooled and field cooled magnetization, **(b)** *M*(H) loop at 5 K after zero field cooling from room temperature (the inset shows the enlarged view of *M*(H) loop indicating symmetric coercive field on positive and negative field-axis) **(c)** *M*(H) loops at 5 K after field-cooling from 350 K in a +3 T field (dark yellow circles) and in a −3 T field (orange circles) **(d)** schematic representation of two possible growth structures, and (e) the thermo-remnant magnetization of [SCO$_{PC}$ (~ 24nm)/SCO$_{BM}$ (~ 2nm)] bi-layer.

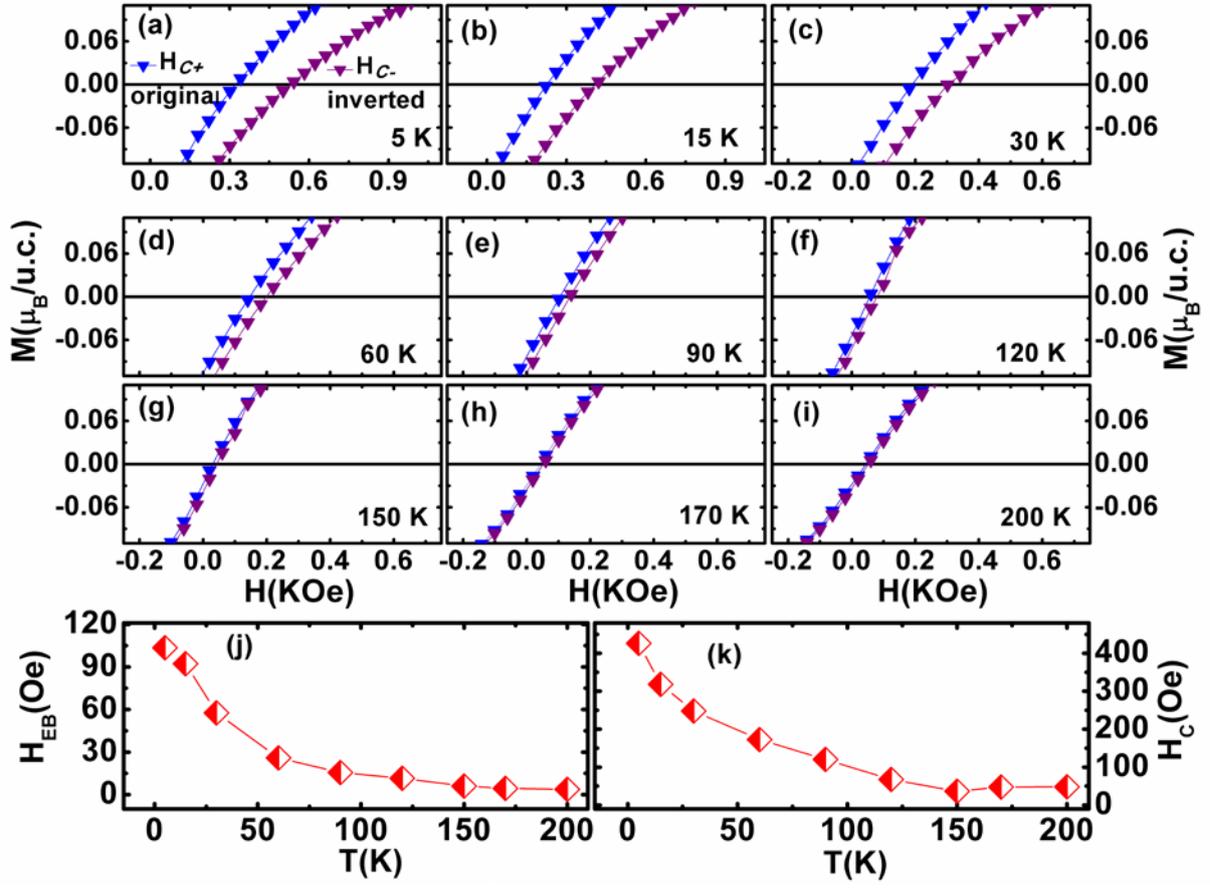

**Figure 3**: Temperature dependence of magnetic exchange bias effect: **(a)-(i)** enlarged view of the original and inverted *M*(H) loops of the [$SCO_{PC}$ (~ 24nm)/$SCO_{BM}$ (~ 2nm)] bi-layer, **(j)** and **(k)** show the estimated $H_{EB}$ and $H_C$ as a function of temperature, respectively.

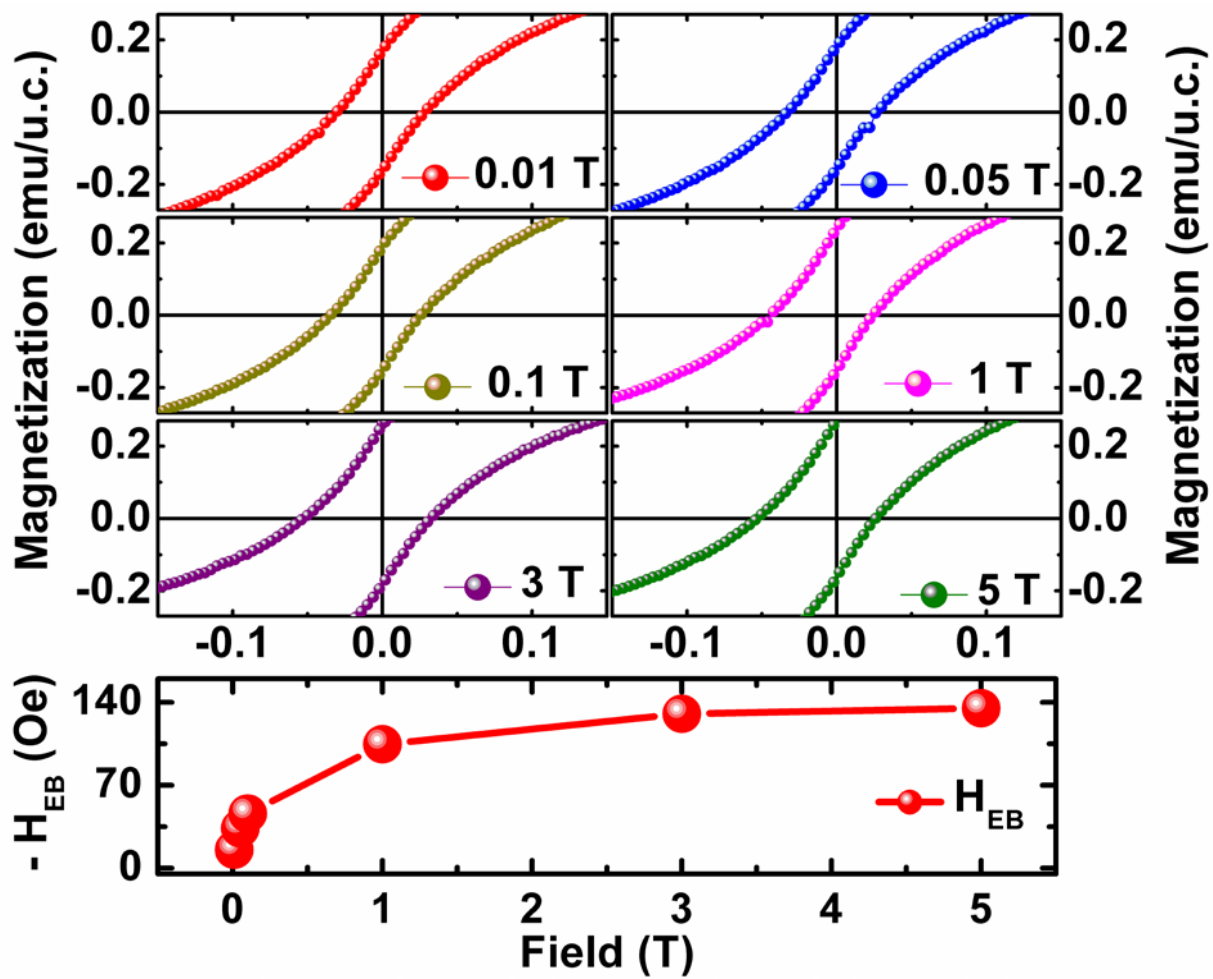

**Figure 4**: *M*(H) loops at 5 K after field-cooling (at various fields) from 350 K (top panel). $H_{EB}$ as a function of field (bottom panel).

**Supporting information for "Unconventionalexchange bias coupling at perovskite/brownmillerite interface in spontaneously stabilized SrCoO$_{3-\delta}$/SrCoO$_{2.5}$ bi-layer"**

B. C. Behera[1], Subhadip Jana[1], Shwetha G. Bhat[2], G. Tripathy[1, 3], P. S. Anil Kumar[2], D. Samal[1,3]

[1]Institute of Physics, Sachivalaya Marg, Bhubaneswar – 751005, India

[2]Department of Physics, Indian Institute of Science Bangalore – 560012, India

[3]Homi Bhabha National Institute, AnushaktiNagar, Mumbai – 400085, India

**1. Structural characterization**

The structural characterization was carried out using high-resolution X-ray diffractometer (Rigaku, Smart Lab). **Figure** S1 shows the $\theta$-$2\theta$ XRD patterns of **(a)** [SCO$_{PC}$ (~ 20 nm)/SCO$_{BM}$ (~ 2 nm)] and **(b)** [SCO$_{PC}$ (~ 24 nm)/SCO$_{BM}$ (~ 2 nm)] bi-layer films. The XRD patterns revealed only noticeable (*00l*) oriented peaks of SCO$_{PC}$, SCO$_{BM}$ indicating highly oriented growth on STO. The out-of-plane lattice parameter "*c*" was estimated to be 3.807 Å and 3.920 Å for SCO$_{PC}$ and SCO$_{BM}$ layers present in the [SCO$_{PC}$ (~ 20 nm)/SCO$_{BM}$ (~ 2 nm)] bilayer. The observed *c*-axis lattice parameter for both SCO$_{PC}$ and SCO$_{BM}$ layers are found to be smaller than their bulk counterpart ($c_{pc}$ = 3.829 Å, and in a pseudo-tetragonal setting, the lattice parameters of SCO$_{BM}$ can be expressed as $c_t$ = 3.936 Å), which indicates both the layers experience an out-of-plane compressive strain on the STO. Similar is the case for [SCO$_{PC}$ (~ 24 nm)/SCO$_{BM}$ (~ 2 nm)] bi-layer in which the *c*-axis lattice parameter turns out to 3.808 Å and 3.920 Å for SCO$_{PC}$ and SCO$_{BM}$ layers respectively.

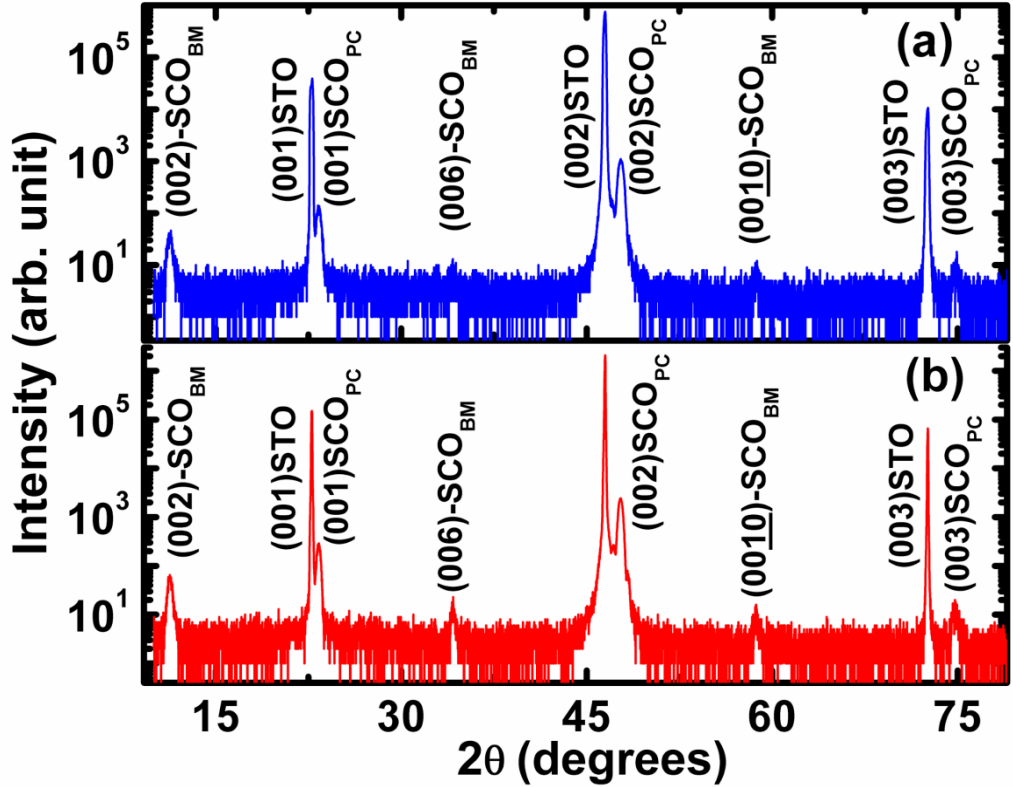

**Figure S1**: *θ-2θ* XRD patterns of **(a)** [$SCO_{PC}$ (~ 20 nm)/$SCO_{BM}$ (~ 2 nm)] and **(b)** [$SCO_{PC}$ (~ 24 nm)/$SCO_{BM}$ (~ 2 nm)] bi-layers grown on the STO.

X-ray reflectivity (XRR) measurement was carried out and the data was fitted using Global fit software of Rigaku to determine thickness, roughness and stacking order sequence in the bi-layers. **Figure** S2 shows the XRR profile of [$SCO_{PC}$ (~ 20 nm)/$SCO_{BM}$ (~ 2 nm)] bi-layer not included in main text. The observed XRR data for the above bi-layer was nicely fitted with [STO/$SCO_{BM}$/$SCO_{PC}$] stacking-order indicating the formation of a natural bi-layer. The possibility for oxygen vacancies in proximity to surface could drive the top layer into BM phase resulting in such natural bi-layers. As mentioned in the main text, the reverse order of stacking such as [STO/$SCO_{BM}$/$SCO_{PC}$] was ruled out as it resulted in an inadequate fit to the XRR data. The root mean square (rms) roughness was found to be ~ 0.43 nm in [$SCO_{PC}$ (~ 20 nm)/$SCO_{BM}$ (~ 2 nm)] bi-layer indicating the layers were stacked in a very smooth fashion.

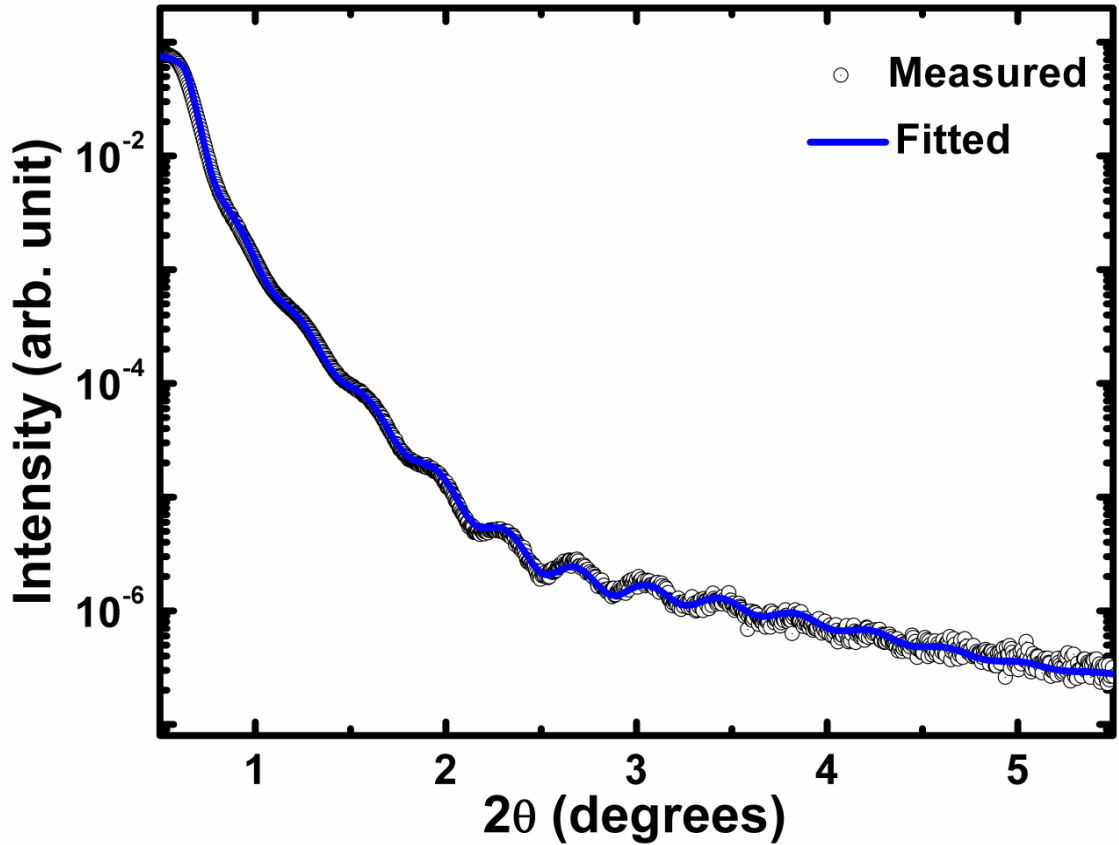

**Figure S2**: XRR profile of [SCO$_{PC}$ (~ 20 nm)/SCO$_{BM}$ (~ 2 nm)] bi-layer on STO.

## 2. M *vs* T magnetic measurement on [SCO$_{PC}$ (~ 20 nm)/SCO$_{BM}$ (~ 2 nm)] bi-layer

Magnetic measurements were carried out using a superconducting quantum interface device based magnetometer (Quantum Design SQUID-VSM). **Figure S3** shows the temperature dependent magnetization *M*(T) (field cooled (FC) and zero-field cooled (ZFC)) of [SCO$_{PC}$ (~ 20 nm)/SCO$_{BM}$ (~ 2 nm)] bi-layer with a field of 100 Oe applied along the in-plane direction of (001) STO. The bi-layer exhibited an onset of ferromagnetic order at $T_C$ ~ 175 K, however finite magnetization was observed above the $T_C$ unlike a conventional FM-paramagnetic (PM) transition. In essence, the characteristic features observed in the *M*(T) plot for [SCO$_{PC}$ (~ 20 nm)/SCO$_{BM}$ (~ 2 nm)] are similar to [SCO$_{PC}$ (~ 24 nm)/SCO$_{BM}$ (~ 2 nm)] bi-layer that is discussed in the main text.

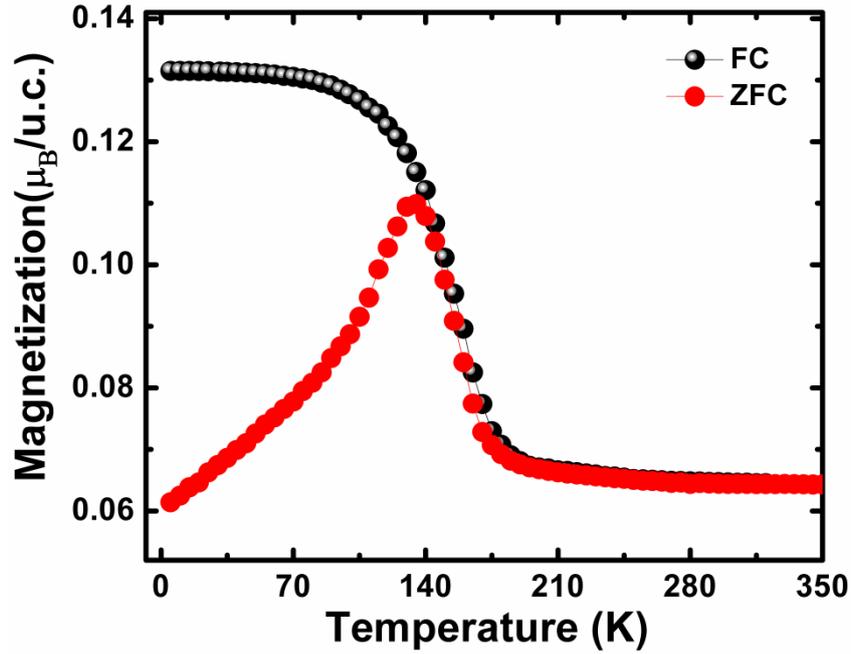

**Figure S3**: ZFC and FC $M$(T) for [SCO$_{PC}$ (~ 20 nm)/SCO$_{BM}$ (~ 2 nm)] bi-layer grown on STO.

## 3. Substrate background correction for the magnetic data

**Figure S4 (a)** shows the virgin $M$(H) plot of [SCO$_{PC}$ (~ 24 nm)/SCO$_{BM}$ (~ 2 nm)] bi-layer, measured at 5 K after the sample is cooled from room temperature to 5 K in the absence of magnetic field. To get rid of STO contribution from the magnetic data and investigate the magnetic signal originating only from the bilayer without any ambiguity, we measured the zero field cooled (ZFC) field dependent magnetization of the STO substrate at 5 K which is shown in **Figure S4 (b)**. Finally, the corrected ZFC $M$(H) plot at 5 K for [SCO$_{PC}$ (~ 24 nm)/SCO$_{BM}$ (~ 2 nm)] bi-layer was obtained (**FigureS4(c)**) after subtracting the STO background from virgin $M$(H) plot.

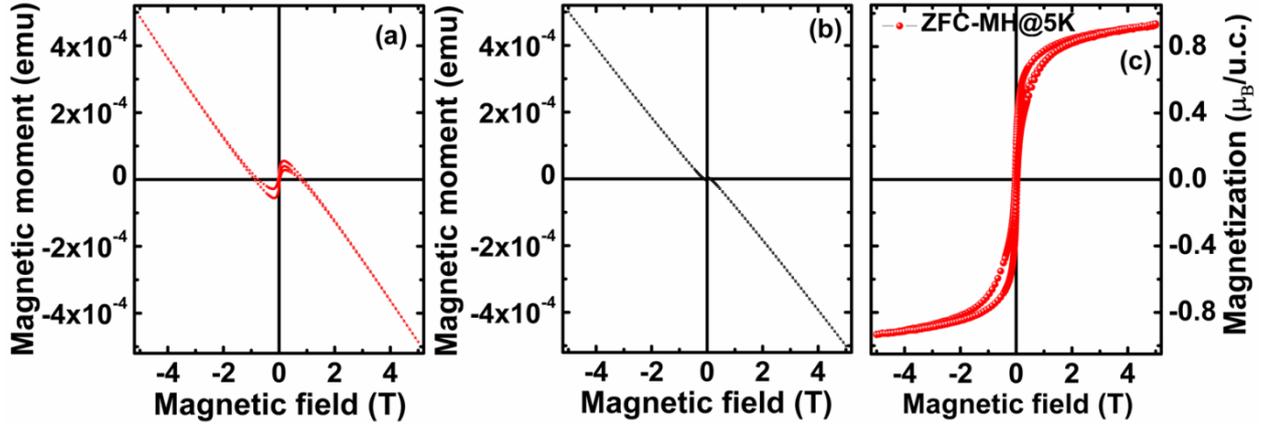

**Figure S4**. **(a)** virgin $M(H)$ loop of the [SCO$_{PC}$ (~ 24 nm)/SCO$_{BM}$ (~ 2 nm)] bi-layer on STO, measured at 5 K after the sample is cooled from room temperature to 5 K in the absence of magnetic field, **(b)** ZFC $M(H)$ plot at 5K for only a diamagnetic STO substrate (on which the bilayers were fabricated) and (c) ZFC $M(H)$ loop of [SCO$_{PC}$ (~ 24 nm)/SCO$_{BM}$ (~ 2 nm)] bi-layer after subtracting the STO background contribution.

## 4. Protocol used for thermo-remnant magnetization (TRM) measurement

In order to rule out the possibility of coupled FM/SG interface resulting in MEBE in (SCO$_{PC}$/SCO$_{BM}$) bi-layer, we measured the thermo remnant magnetization (TRM) using the following protocol. The [STO/SCO$_{PC}$ (~ 24 nm)/SCO$_{BM}$ (~ 2 nm)] bi-layer was cooled down from room temperature to measuring temperature ($T_M$) 100 K under 500 Oe applied field. When the $T_M$ was reached, the applied field was kept for 300 s. After that, the field was removed and the magnetization was measured as a function of time. The reference time corresponded to the time at which applied field was set to zero. It was evident from the data shown in main text **Figure2 (e)** that the magnetization remains constant over four decades of time, ruling out any spin glass character. Had there been any spin glass character, it was natural to expect time dependent changes in the magnetization.

## 5. Residual field of the magnet

To minimize the residual magnetic field in the SQUID magnetometer, the magnetic field (H) was set to zero in an oscillation mode before measurements. The residual field was further calibrated by a reference Pd sample that shows a negligible value around - 1.0 Oe (see **Figure S5**).

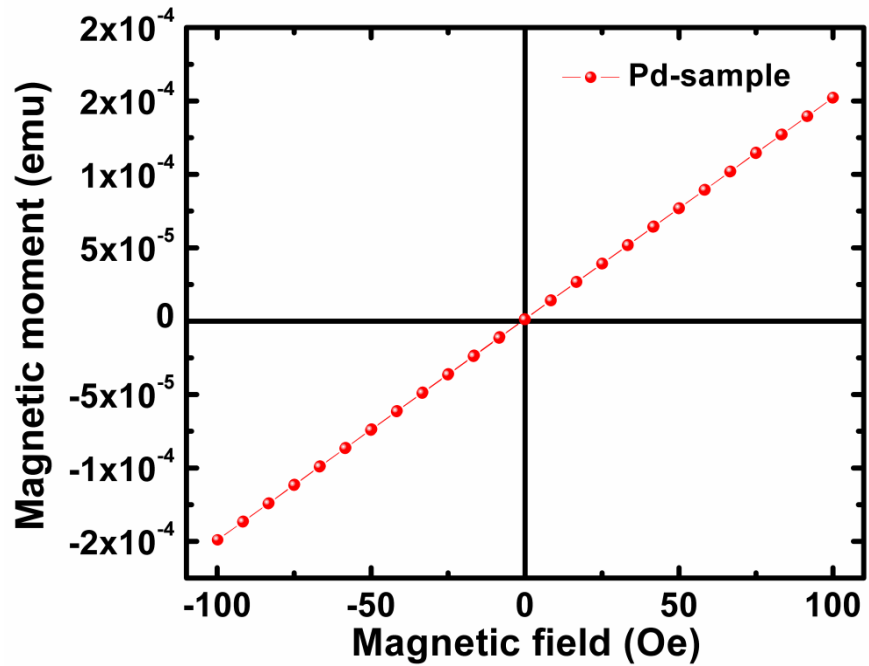

**Figure S5**. Zero field cooled magnetic hysteresis loop of the Pd sample measured at 300 K.

## 6. Magnetic hysteresis loop for [SCO$_{PC}$ (~ 24 nm)/SCO$_{BM}$ (~ 2 nm)] bi-layer at various temperatures:

Magnetic exchange bias effect (MEBE) reflects in the asymmetric shift of the magnetic hysteresis loop along the field-axis coupled with the enhancement of coercivity induced by unidirectional exchange anisotropy-field at a FM/AF interface. In the main text we used so called "inversion" method to illustrate unambiguously the temperature dependent MEBE for [SCO$_{PC}$ (~ 24 nm)/SCO$_{BM}$ (~ 2 nm)] bi-layer. The "inverted loop" is produced by the "inversion" method in which magnetization and the magnetic field of the original loop are multiplied by -1. Here, we present the temperature evolution in the asymmetry of the non-inverted $M$(H) loop for [SCO$_{PC}$ (~ 24 nm)/SCO$_{BM}$ (~ 2 nm)] bi-layer (from which the "inverted loop" was made) for clarity (see **Figure S6**). Remarkably, an S-kind of hysteresis with a finite $H_C$ was observed even at 200 K which is above the $T_C$ ~ 175 K of SCO$_{PC}$ layer. Essential procedures and precautions such as setting the magnetic field to zero in an oscillation mode and considering the tiny residual field (~1 Oe) from Pd sample calibration was taken into account to estimate precisely the exact values of coercive field ($H_C$) and magnetic exchange bias field ($H_{EB}$) as shown in **Figures 3 (j)-(k).**

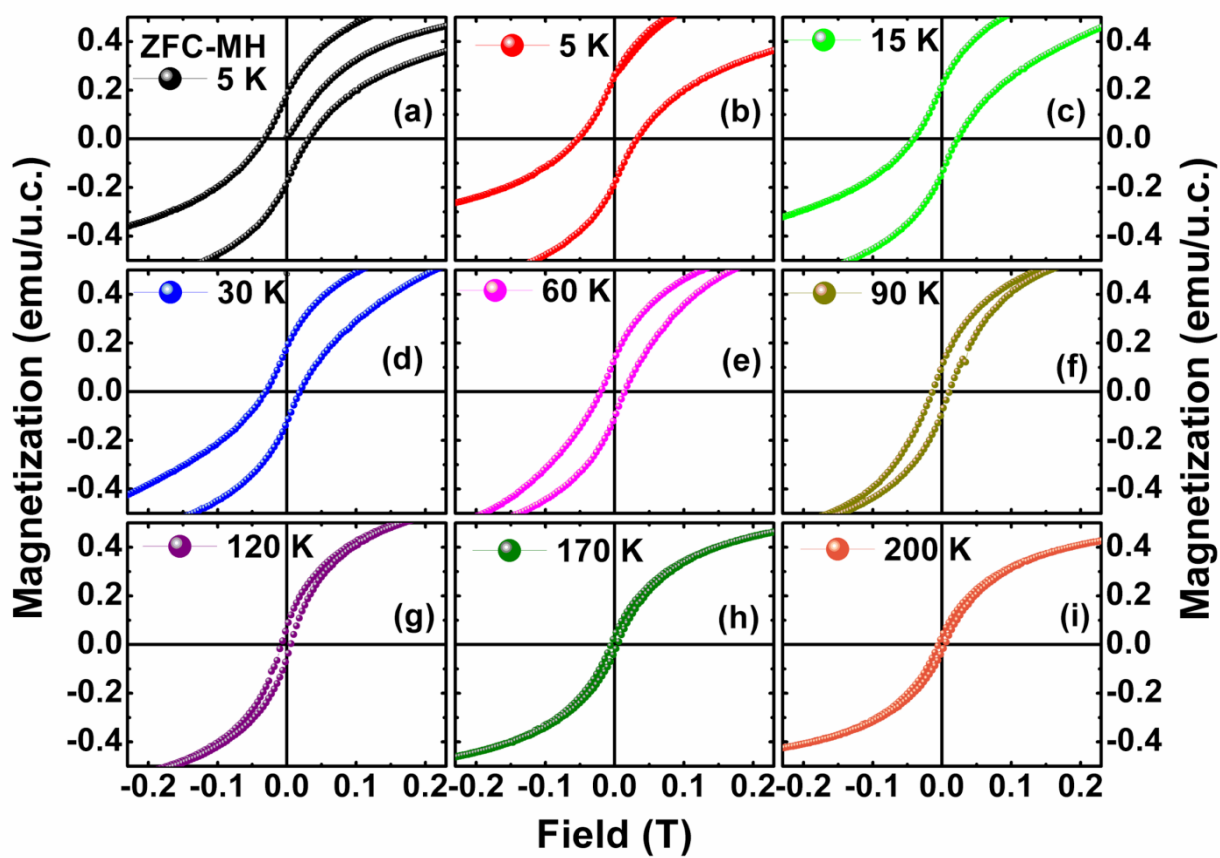

**Figure S6**. **(a)** ZFC M-H loop of [SCO$_{PC}$(~ 24 nm)/SCO$_{BM}$ (~ 2 nm)]bi-layer at 5K, **(b) – (i)** FC magnetic hysteresis loop of[SCO$_{PC}$(~ 24 nm)/SCO$_{BM}$ (~ 2 nm)]bi-layer measured at various temperatures (cooling was done from 350 K to the measurement temperature in the presence of applied magnetic field 3 Tesla).

## 7. Temperature dependent $H_{EB}$ and $H_C$ for [SCO$_{PC}$ (~ 20 nm)/SCO$_{BM}$ (~ 2 nm)] bi-layer

We have also investigated the MEBE on [SCO$_{PC}$ (~ 20 nm)/SCO$_{BM}$ (~ 2 nm)] bi-layer. Temperature dependent exchange bias field ($H_{EB}$) and the coercive field ($H_C$) of the above bi-layer are presented in the **Figure S7**. Here, we also observed that the $H_{EB}$ gradually decreases with increasing temperature and falls to zero in the vicinity of $T_C$ ~ 175 K (see **Figure S7 (a)**). Remarkably, the non-zero $H_C$ is preserved above $T_C$ (see **Figure S7(b)**) which is very similar to the observation made in the case of [SCO$_{PC}$ (~ 24 nm)/SCO$_{BM}$ (~ 2 nm)] bi-layer as illustrated in the main text.

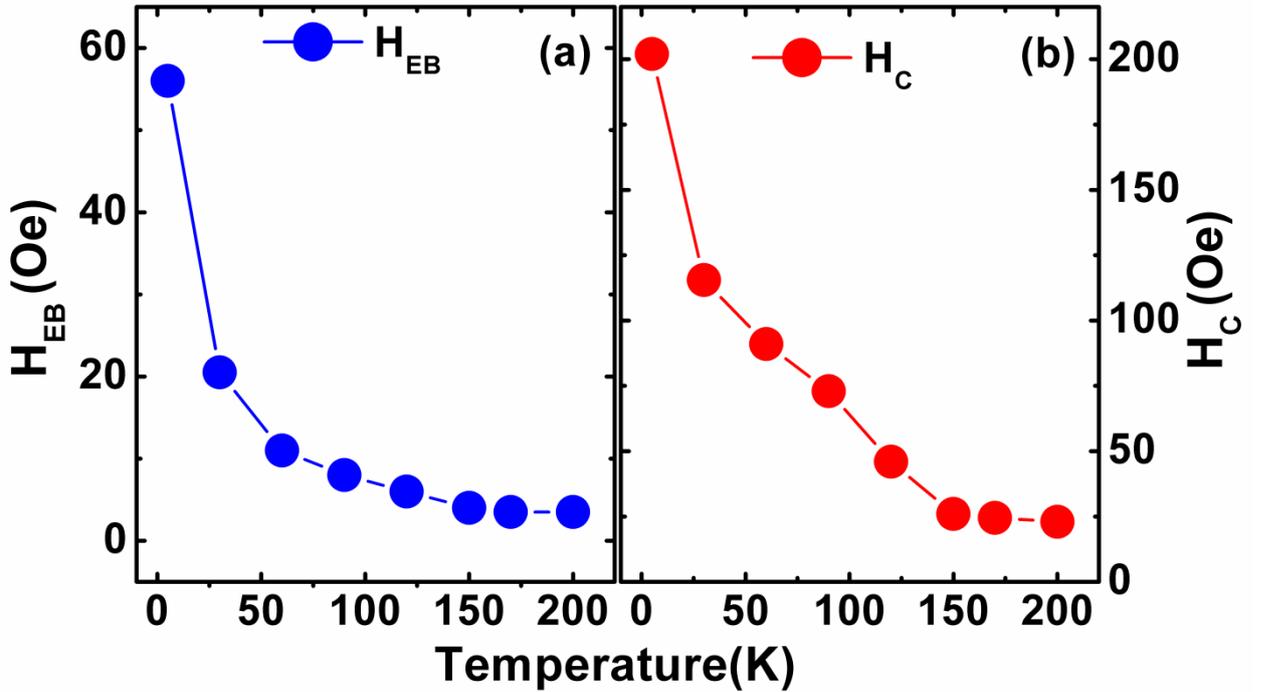

**Figure S7**. Temperature dependence of the **(a)** $H_{EB}$ and **(b)** $H_C$ of [SCO$_{PC}$ (~ 20 nm)/SCO$_{BM}$ (~ 2 nm)] bi-layer grown on STO.

## 8. Results from Monte Carlo simulations:

To develop a better theoretical understanding of this effect, we performed Monte Carlo simulations of a simple two dimensional Ising model of the FM-AF bilayer system with non-flat interface at atomic scale (**Figure S8**). The interfacial coupling between the FM and AF spins is chosen to be ferromagnetic. In **Figure S8**, we show the magnetic hysteresis loop data from the simulations for $T>T_C$ and schematically show the microscopic spin configurations (in the time averaged sense) at a number of representative points on the hysteresis loop. While a flat interface between the FM and AF layers reproduces the non-zero $H_C$ and vanishing $H_{EB}$ for $T>T_C$, it fails to display a non-zero $H_{EB}$ for $T<T_C$, as observed in **Figure 3 (j)**. On the other hand, an interface which is non-flat (only on the atomic scale, as mimicked by single lattice spacing width of the interface in **Figure S8**), gives results which are in fairly good qualitative agreement with the data observed for $H_{EB}$ and $H_C$ (**Figures 3 (j)and (k)**) for $T<T_C$ as well as $T>T_C$.

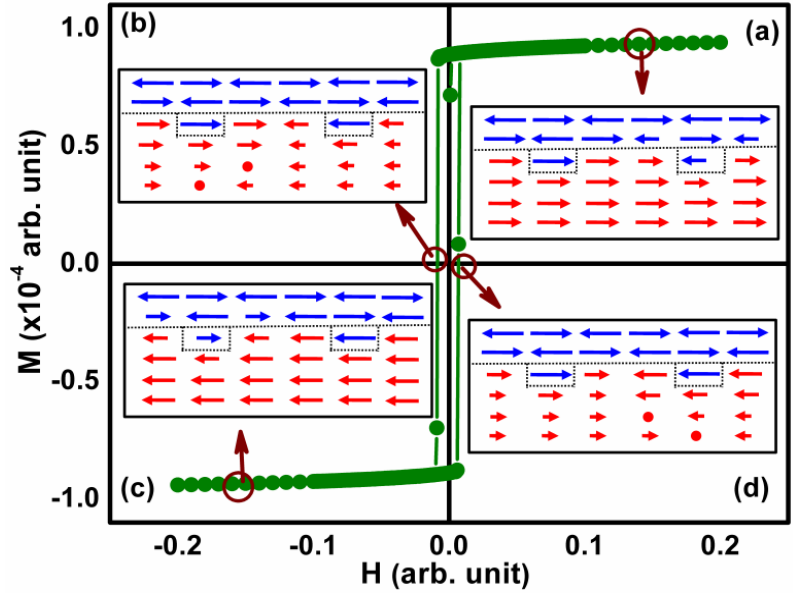

**Figure S8**. Magnetic hysteresis loop (in green) from Monte Carlo simulation of an Ising model of the FM(SCO$_{PC}$)/AF(SCO$_{BM}$) bi-layer above the Curie temperature of FM(SCO$_{PC}$) layer. The insets show schematic microscopic spin configurations at different points of the hysteresis loop. The red and blue arrows represent the spin magnetic moments of FM(SCO$_{PC}$) layer and AF(SCO$_{BM}$) layer, respectively. The lengths of the arrows are approximately related to the ensemble averaged value of the corresponding magnetic moment (a dot represents vanishing value). For the simulations we have chosen $J_F/k_BT= 0.53$, $J_{AF}/k_B= -1.5$ and $J_I/k_BT = 0.8$ (the critical value for 2D Ising model is 0.55). The system size in the simulation is 100 x 100 spins with 97 FM layers and 2 AF layers, apart from the interfacial layer. The locations of the AF spins in the interfacial layer are chosen randomly. We believe the effective domain flipping (e.g. set of spins surrounding the second AF spin in the interface layer flipping between insets **(a)** and **(b)** ) gives rise to the hysteresis seen in the paramagnetic phase of the AF layers.